\documentclass{PoS}
\usepackage{url}

\title{Observations of the extreme runaway HD\,271791: nucleosynthesis in a core collapse supernova}

\ShortTitle{The extreme runaway HD\,271791: nucleosynthesis in a core-collapse supernova}

\author{\speaker{Veronika Schaffenroth}\\
        Institute for Astro- and Particle Physics, University of Innsbruck, Technikerstr. 25/8, A-6020 Innsbruck, Austria {\bf and}\\        
        Dr.~Karl Remeis--Observatory \& ECAP, Astronomical Institute,
        Friedrich-Alexander University Erlangen-Nuremberg, Sternwartstr. 7, D-96049 Bamberg, Germany\\
        E-mail: \email{veronika.schaffenroth@uibk.ac.at}}

\author{Norbert Przybilla\\
        Institute for Astro- and Particle Physics, University of Innsbruck, Technikerstr. 25/8, A-6020 Innsbruck, Austria}
\author{Keith Butler\\
	    Munich Observatory, Ludwig-Maximilians University, Scheinerstr. 1, D-86179 Munich, Germany}   
\author{Andreas Irrgang, Ulrich Heber\\
       	 Dr.~Karl Remeis--Observatory \& ECAP, Astronomical Institute, Friedrich-Alexander University Erlangen-Nuremberg, Sternwartstr. 7, D-96049 Bamberg, Germany
       	         }   

\abstract{Some young, massive stars can be found in the Galactic halo. As star formation does not occur in the halo, they must have been formed in the disk and been ejected shortly afterwards. One explanation is a supernova in a tight binary system. The companion is ejected and becomes a runaway star. HD\,271791 is the kinematically most extreme runaway star known (galacic restframe velocity $725 \pm 195\, \rm km\,s^{-1}$ even greater than the Galactic escape velocity). Moreover, an analysis of the optical spectrum showed an enhancement of the $\alpha$-elements. This indicates an origin in a supernova. As such high velocities are not reached in classical binary supernova scenarii, a very massive but compact primary, probably of Wolf-Rayet type is required. The star is a perfect candidate for studying nucleosynthesis in a core-collapse supernova because of the contamination of its surface layers with supernova ejecta of its former very massive primary. The goal of this project is to determine the abundances of a large number of elements from the $\alpha$-process, the iron group, and heavier elements by a quantitative spectral analysis from the optical and the UV with detailed stellar atmosphere models that account for deviations from the local thermal equilibrium (NLTE). We intent to verify whether core-collapse supernova are a site of r-process element production. Here, we state the current status of the project.
 }

\FullConference{XIII Nuclei in the Cosmos,\\
		7-11 July, 2014\\
		Debrecen, Hungary }

\begin{document}

\section{Runaway stars}
Young, massive stars are usually found close to the Galactic plane, preferentially in open clusters and associations. Some of them, however, are observed at high Galactic latitudes far away from star-forming regions. Since no gas clouds are known in the halo that have a sufficient density to form massive stars, these stars must have formed in the Galactic disc, and afterwards migrated outwards ('run-away B stars'). They are thought to have been ejected from their place of birth and accelerated to high velocity by dynamical processes either during the initial dynamical relaxation of a star cluster (Poveda et al. 1967), or in binary interactions inside star clusters (Leonard \& Duncan 1988), or by means of a binary supernova explosion (Blaauw 1961). 

In the latter scenario the companion is ejected with about the orbital velocity and becomes a runaway star. To distinguish the supernova scenario from the other scenarios an abundance study can be used, as the atmosphere of the close companion, afterwards the runaway star, is polluted by the supernova ejecta. Therefore, an enrichment of the elements existing in the supernova ejecta is expected. Such stars are perfect candidates for studying nucleosynthesis in a core-collapse supernova, which is theoretically under debate right now.
\begin{figure}[h]
\begin{minipage}{0.5\linewidth}
\includegraphics[width=1.0\linewidth]{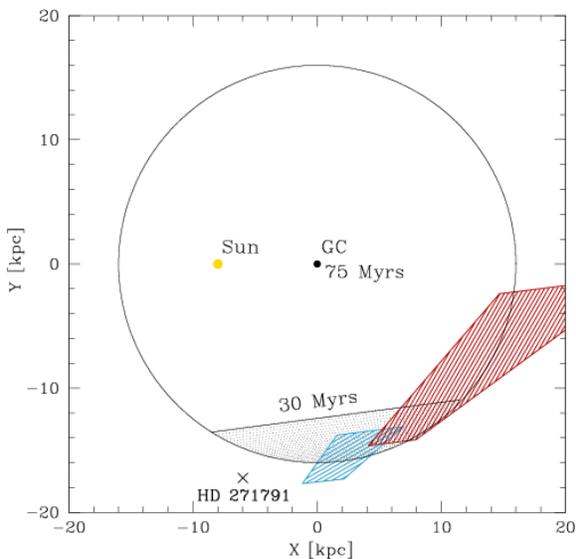}
\caption{Regions of origin for HD 271791 in the Galactic plane calculated by varying the
proper-motion components within their measurement errors. The position of HD 271791 projected to the Galactic plane is marked. The red area is derived from our proper-motion measurement, whereas the blue area follows if proper motions from the UCAC2 catalog are used.}
\end{minipage}\hfill\hspace{0.5cm}
\begin{minipage}{0.45\linewidth}
HD\,271791 is the kinematically most extreme runaway star known (galactic restframe velocity $725 \pm 195\, \rm km\,s^{-1}$, even greater than the Galactic escape velocity). Such velocities were believed only to be reached, if a star is ejected by the supermassive black hole in the centre of the Galaxy (Hills 1988). However, HD\,271791 is much younger than the flight-time from the centre of the Galaxy to its current position. Moreover, a reconstruction of the orbit shows that HD\,271791 comes from the outskirts of the Galaxy (see Fig. 1). This rules out the supermassive black hole scenario. Unfortunately it is not possible to observe clusters in the region of the birthplace of HD\,271791. Therefore, the ejection due to dynamical interactions in a cluster cannot be ruled out. However, an abundance study can be used to investigate the supernova scenario. 
\end{minipage}
\end{figure}

\section{Observations}
To check for enrichment in the abundances of HD\,271791 we want to perform an abundance study in the optical and the UV. 
Therefore, we obtained high-S/N optical ESO-VLT UVES spectra from 3000 to 10\,000 \AA\, with a resolving power R = 30\,000.
In the UV we got HST/COS spectra from 1150 to 1800 \AA\,with a resolution of 16\,000-21\,000 and HST/STIS from 1600 to 3100 \AA\, with a resolution of 30\,000. Hence, we are covering the whole wavelength range from 1150 to 10\.000 \AA.

\section{Method}
In this section we want to describe the method used in our project shortly.
As a first step the parameters of the star and the elemental abundances are determined from the $\alpha$- elements with NLTE models. Before we can use the observations in the UV for deriving abundances, synthetic spectra in the UV have to be developed first, as until now we analysed only optical spectra. As the lines of the runaway star are very broad due to a high rotational velocity, they are not suitable to test the synthetic spectra. Therefore, we use very bright B stars with similar parameters and slow rotation ($\iota$ Her, $\gamma$ Peg, HR1840) to develop NLTE models in the UV. 

For the calculation of synthetic spectra we use a hybrid NLTE approach. Model atmospheres are calculated in LTE with {\sc atlas} (Kurucz 1993). This allows the use more complex model atoms.
The radiative transfer and statistical equilibrium is solved by {\sc detail} (Butler \& Giddings 1985). Moreover, it calculates NLTE populations. The synthetic spectra are finally calculated with {\sc surface}.

Based on the atmospheric parameter and abundances determined in the optical, we derive elemental abundances of the iron group and heavier elements in the UV. To avoid systematic effects and to derive the abundances only from the polluting supernova ejecta the analysis is done differentially with respect to a representative B-star sample (Irrgang et al. 2014, in prep.). The derived enrichments in the elemental abundances are finally compared to theoretical calculations of yields in core-collapse supernovae by Nomoto et al. (2006).
\section{Analysis}
\subsection{Optical}
We analysed the optical spectrum of HD\,271791 by fitting NLTE synthetic spectra to the whole spectrum. More details on the spectrum synthesis can be found in Sect 4.2. The derived parameters can be found in Table 1.
  \begin{table}[h]
  \caption{Parameter of HD\,271791 derived by the analysis of the optical UVES spectrum}
  \begin{center}
  \begin{tabular}{lcclcc}
  \hline\hline
  Parameter&Value&Unit&Parameter&Value&Unit\\\hline
  $T_{\rm eff}$&$18700\pm50$&K&$v \sin i$&$128\pm1$&$\rm km\,s^{-1}$\\
  $\log g $&$3.15\pm0.02$&cgs&$\zeta$&$22\pm 4$&$\rm km\,s^{-1}$\\
  $\xi$&$5.9\pm0.1$&$\rm km\,s^{-1}$&$v_{\rm rad}$&$443\pm1$&$\rm km\,s^{-1}$\\
  \hline\hline
  \end{tabular}
    \end{center}
  \end{table}
\begin{figure}[h]
\begin{minipage}{0.5\linewidth}
\includegraphics[angle=-90,width=1.0\linewidth]{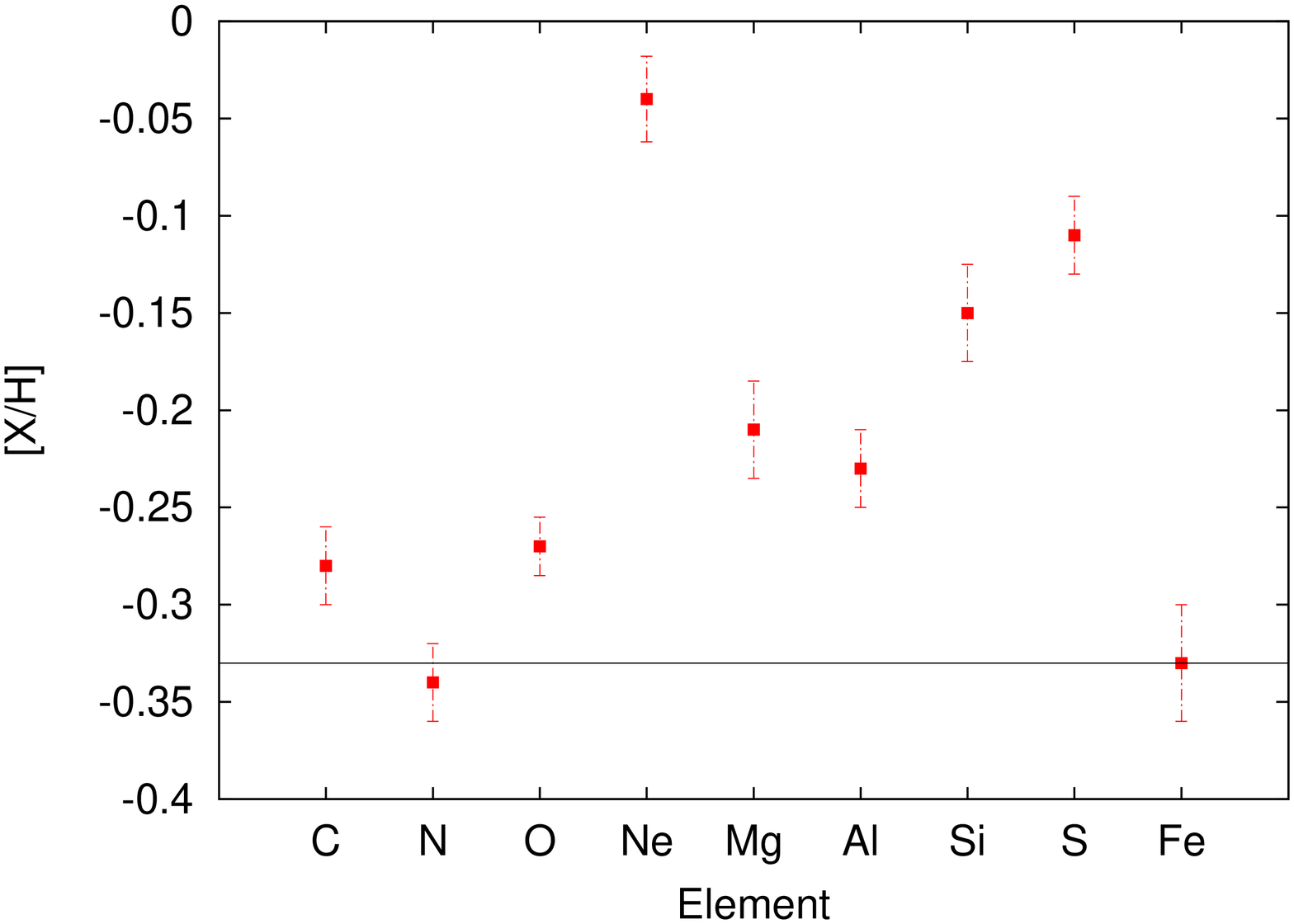}
\end{minipage}
\begin{minipage}{0.5\linewidth}
\includegraphics[width=1.0\linewidth]{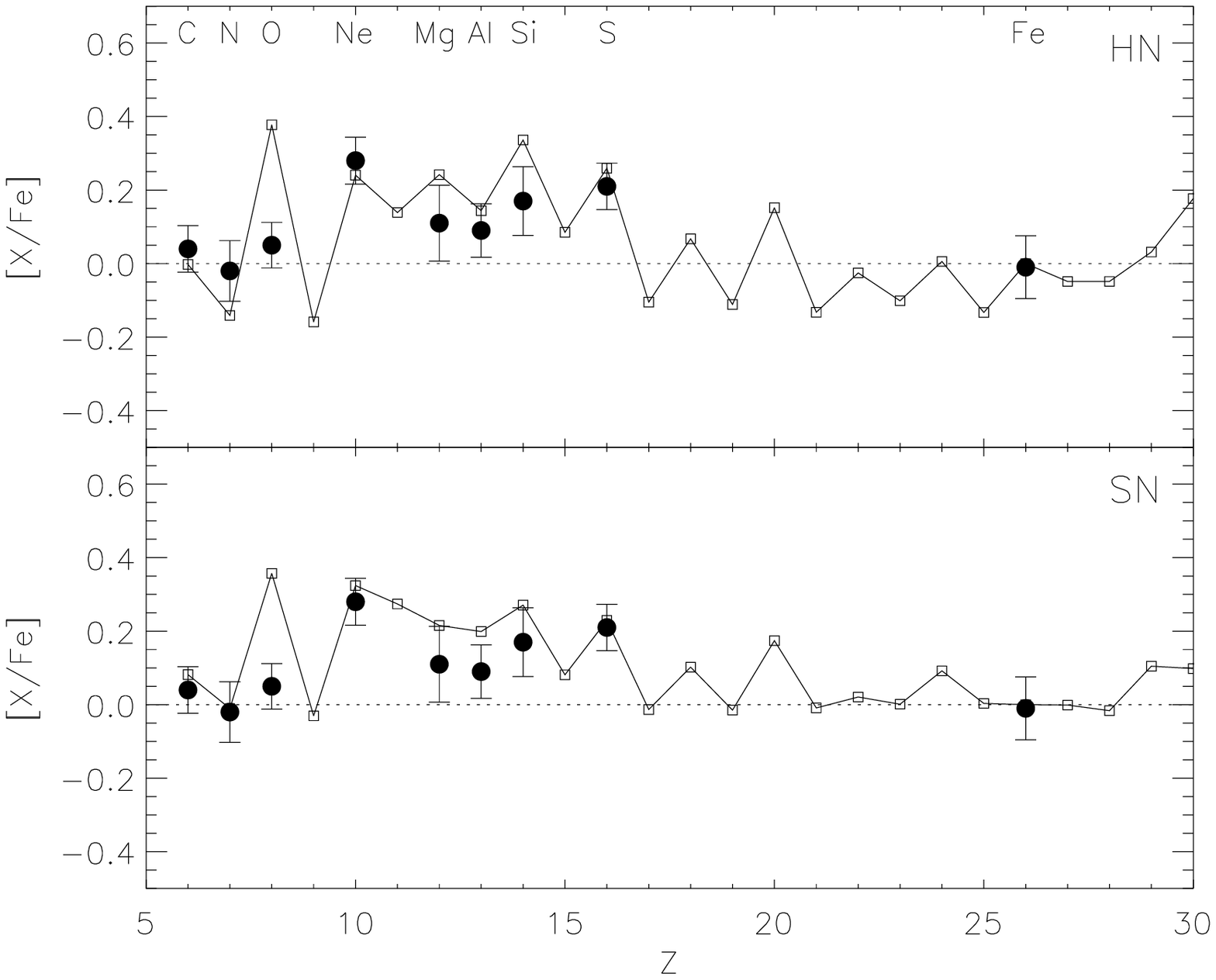}
\end{minipage}
\caption{Abundances of HD\,271791 determined from the optical spectrum. The left figure shows the abundances relative to a representative B-star sample (Irrgang et al. 2014, in prep.). The baseline metallicity of HD 271791, [Fe/H], is marked by the solid line. The right figure shows the abundances normalised to iron compared to hypernova/supernova (HN/SN) yields by Nomoto et al. (2006)}
\label{optical}
\end{figure}
In Fig. 2 the abundances in comparison to the representative B star sample by Irrgang et al. (in prep.) of 63 nearby B stars are shown. As it is not expected that the supernova ejecta polluting the atmosphere of the runaway contains iron, we use iron as a baseline. It is obvious that the iron abundance is about 0.3 dex lower than the iron abundance of the B star sample. However, that is not surprising, as HD\,271791 originates from the outskirts of the Galaxy. Lower metallicities are observed there. Moreover, several of the $\alpha$-elements e.g. Ne, Mg, Si and S are enhanced compared to the abundances expected because of the metallicity of HD\,271791. Figure 2 shows on the right side the comparison of the elemental abundances normalised to iron with hypernova/supernova yields by Nomoto et al (2006). A qualitative agreement between theory and observation is reached for both the hypernova and the supernova yields. Only the oxygen abundance is smaller than would be expected by theory. However, that can be explained because of the use of integrated yields. A homogeneous distribution in the supernova ejecta is not expected. Therefore, simulations of the supernova explosion and the accretion of supernova ejecta on the runaway are needed to get are more realistic supernova yields. The enrichment in the $\alpha$-elements that is observed indicates an ejection of HD\,271791 by a supernova explosion in a very tight system. To explain the extreme velocity a very massive but compact primary, probably of Wolf-Rayet type is required.

\subsection{UV}
\begin{figure}[h]
\includegraphics[width=0.41\linewidth]{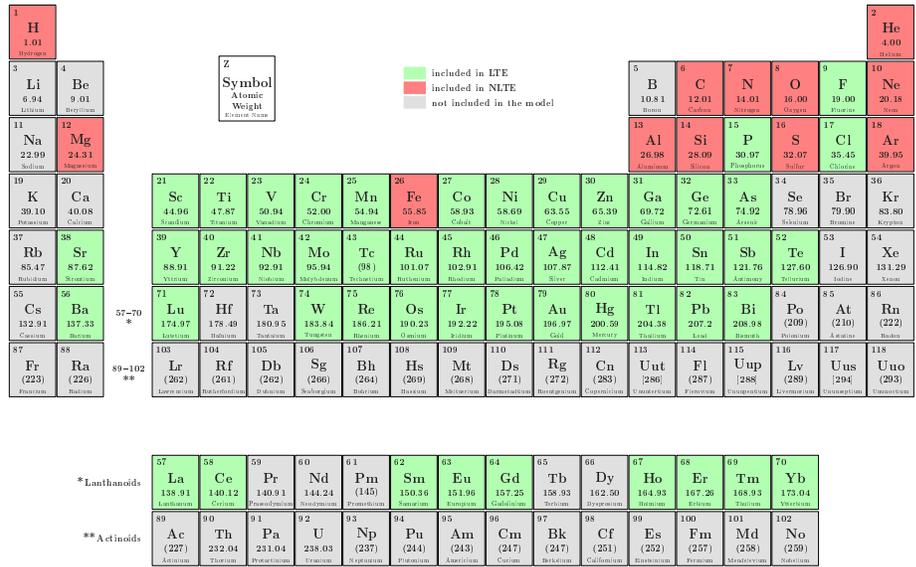}
\caption{Periodic table showing all elements implemented in our spectrum synthesis. The red marked elements are available in non-thermal equilibrium (NLTE), the green marked ones only in LTE.}
\label{pt}
\end{figure}
For the development of synthetic spectra in the UV it was first necessary to implement lists of lines in the UV from all desired elements using all available atomic data into the spectrum synthesis. The atomic data was mostly taken from NIST \footnote{\url{http://www.nist.gov/pml/data/asd.cfm}}, Kurucz\footnote{\url{http://kurucz.harvard.edu/}}, Iron Project (Pradhan 2000) and Morton (2000). Figure 3 shows, which elements are now available in our synthetic spectra. The limiting factor of this project is missing atomic data. For many atoms only atomic data for neutral or single-ionized atoms were available. Because of the high effective temperature of HD\,271791, however, only lines of multiple-ionized atoms are visible.

\begin{figure}[h]
\includegraphics[angle=-90,width=0.705\linewidth]{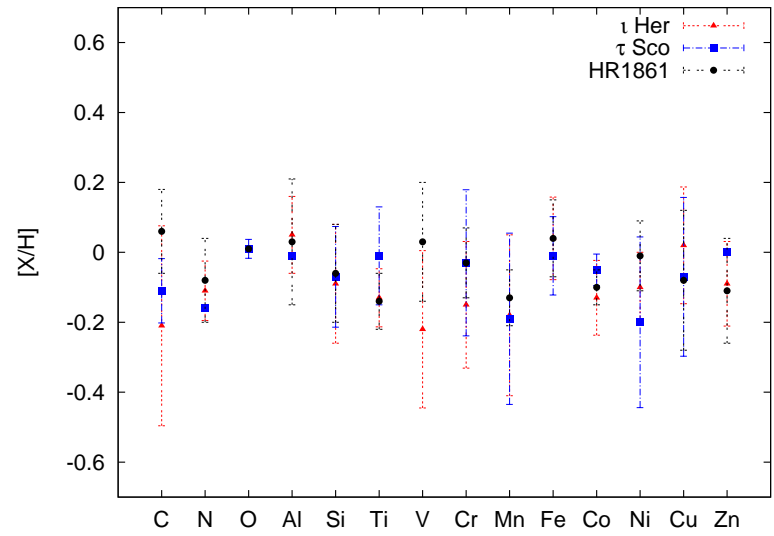}
\caption{Abundances for three comparison stars of different temperatures determined from the UV spectrum relative to solar values. The errors are given by the standard deviation of the abundances determined from single lines.}
\label{abun_UV}
\end{figure}
\begin{figure}[h]
\includegraphics[width=1.1\linewidth]{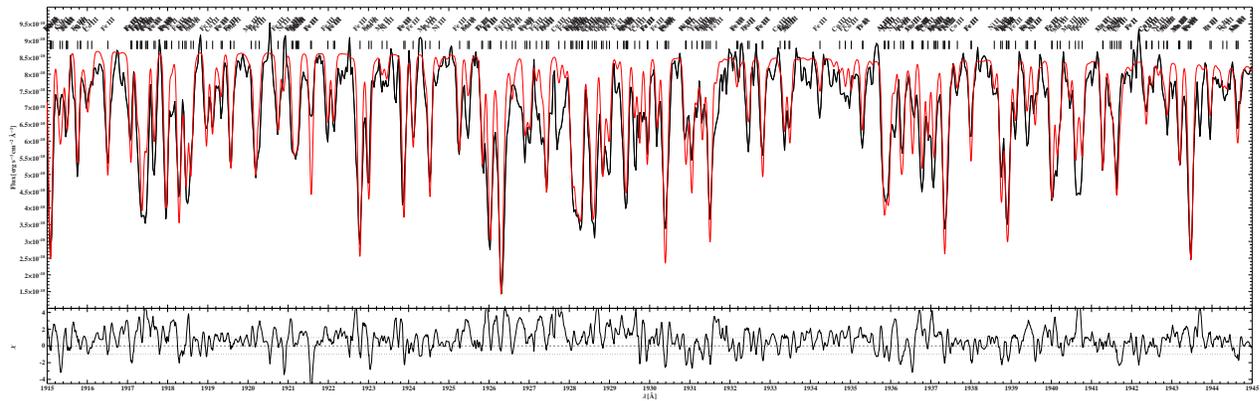}
\caption{Comparison of the observed UV spectrum of $\iota$Her to a synthetic spectrum calculated with abundances from the optical analysis, if available, and solar abundances for the rest of the elements. The lower figure shows the residuals.}
\label{UV_spec}
\end{figure}
To test our spectrum synthesis we derived the abundances of several elements in the UV from single unblended lines. In Fig. 4 the results for three different stars are displayed relative to the abundances in the optical or solar values, if no lines of the element are visible in the optical. The abundances in the UV match the expected values within the error for all three stars. The error is thereby given by the standard deviation of the abundances from the single lines. Figure 5. shows a comparison of a synthetic spectrum calculated with abundances from the optical or solar abundances with an HST/STIS spectrum of $\iota$Her. There are still some lines missing, but most of the lines are fitting quite well. Therefore, we are confident that our spectrum synthesis in the UV is working and can be used to derive reliable abundances.
\section{Outlook}
In the next step we now also want to derive the abundances of the heavier elements from blended lines. Afterwards we want to determine the abundances of the rapidly rotating HD\,271791 and compare them with abundances of the comparison stars to derive the abundances of the iron group and heavier elements in the supernova ejecta. For the future we want to build NLTE model atoms for the iron group elements to minimise NLTE effects, for more accurate abundances. To study nucleosynthesis in a core-collapse SN further, we also want to apply for observing time with HST and analyse more candidate runaway stars from the SN scenario.

\end{document}